\documentstyle[12pt]{article}

\newcommand{\be}{\begin{equation}}
\newcommand{\ee}{\end{equation}}
\newcommand{\bea}{\begin{eqnarray}}
\newcommand{\eea}{\end{eqnarray}}

\def\ep{\epsilon}
\def\lm{\lambda}

\def\ln{{\rm ln}}
\def\ep{\epsilon}
\begin{document}
\setlength{\baselineskip}{0.7cm}

\begin{titlepage}
\null
\begin{flushright}
hep-ph/0209009\\
UT-02-49\\
September, 2002
\end{flushright}
\vskip 1cm

\begin{center}
{\Large\bf 
Decoupling and lepton flavor violation
 in extra dimensional theory}

\lineskip .75em
\vskip 1.5cm

\normalsize
{\large Naoyuki Haba$^{a}$},
{\large Nobuhito Maru$^{b}$}, and
{\large Noboru Nakamura$^{c}$}

\vspace{1cm}

{\it $^{a}$Faculty of Engineering, Mie University, Tsu, 
Mie, 514-8507, JAPAN} \\
{\it $^{b}$Department of Physics, University of Tokyo, 
Tokyo 113-0033, JAPAN} \\
{\it $^{c}$Department of Physics, Nagoya University, 
Nagoya, 464-8602, JAPAN} \\

\vspace*{10mm}

{\bf Abstract}\\[5mm]
{\parbox{13cm}{\hspace{5mm}
%

We discuss  the fermion mass hierarchy and the flavor mixings 
 in the fat brane scenario of five dimensional SUSY theory. The decoupling solution of the sfermion mass spectrum 
 can be realized by introducing the vector-like mirror fields 
 in an extra dimension. 
In this scenario, both the left- and right-handed sleptons 
 can have sizable flavor mixings. We point out that this sizable flavor mixings  can induce the suitable magnitude of the muon anomalous magnetic 
 moment ($g_\mu -2$) 
 within the experimental bounds of lepton flavor violating processes. 
}}

\end{center}

\end{titlepage}


\section{Introduction}
%

Utilizing extra dimensions has shed new insights into 
 various phenomenological aspects of the physics 
 in four dimensions. 
Antoniadis \cite{string} 
proposed the possibility that part
of the standard model particles live in TeV extra dimensions, in
connection to the problem of supersymmetry breaking.
One of the advantages of theories with extra dimensions 
 is that small parameters in the theory can be naturally 
 obtained due to the locality not the symmetry.  
In particular, 
 Arkani-Hamed and Schmaltz \cite{AS} 
 have proposed an interesting mechanism, 
 which is referred to as ``fat brane scenario", 
 in which small parameters are obtained 
 by a small overlap of wave functions, 
 even if the parameters in a fundamental theory are of order unity. 
This mechanism has been applied to various phenomenological issues so 
 far, 
 such as the fermion mass hierarchy 
 \cite{MS,KT,Branco,Branco2,KT2,KY2,HM3,Jsantiago}, 
 the doublet-triplet splitting \cite{KY,Maru,HM}, 
 and the sfermion mass generation \cite{KT,MSSS,KT2,HM2,HM3}.

In the previous paper, 
 two of the present authors (N.H. and N.M.) have discussed 
 the fermion mass hierarchy and the flavor mixings 
 in the fat brane scenario of a five dimensional supersymmetric 
 (SUSY) theory \cite{HM3}. 
In our set up, 
 the matter lives in the bulk, its zero mode 
 wave functions are Gaussian and are localized at different points 
 in extra dimensions. 
On the other hand, Higgs fields are localized on a brane.
The fermion mass hierarchy is determined by the values at a brane 
 where Higgs fields are localized. 
Various types of the matter configurations were found, 
 which yield the fermion mass matrices consistent 
 with experimental data.

As for the sfermion mass spectrum,  
 they are generated by the overlap 
 between the wave functions of matter fields and 
 the chiral superfield with nonzero vacuum expectation value (VEV) 
 of the F-component localized on the SUSY breaking brane. 
In Ref.~\cite{HM2}, 
 we have proposed that if SUSY breaking brane is located 
 between the 1st and the 2nd generations, the sfermion mass 
 spectrum becomes the decoupling solution 
 (sometime called as ``effective SUSY'') \cite{effsusy}. 
In this solution, the squarks and sleptons in the 1st and 2nd 
 generations are heavy enough so that their contributions to the 
 FCNC or CP violating processes are sufficiently suppressed.\footnote{  
On the other hand, the gauginos, higgsinos and the sfermions in 
the 3rd generation are appropriately light to satisfy the 
naturalness condition on the Higgs boson mass.}

On the other hand, 
 the recent experiments again suggest the discrepancy of 
 the muon anomalous magnetic moment from the 
 standard model (SM) prediction 
 \cite{recentg-2}. 
One of the uncertainties of the SM predictions is coming from 
 the evaluation of the hadronic vacuum polarization. 
Davier et al. \cite{Davier} made careful analysis of this contribution 
 (one based on the cross section of $e^+ e^-$ to hadrons and 
 the other one based on $\tau$ decay.). 
The result based on $e^+ e^-$ cross section suggests that 
 the SM prediction is about 3$\sigma$ deviation from the experimental data, 
 while the result based on $\tau$ decay shows that the SM prediction 
 is consistent with the experimental data. 
The orogin of this difference has not been clarified. 
In this paper, we take the $e^+ e^-$-based result by Davier et al. 
 for the hadronic contributions. 
It has been shown in Ref.\cite{CHH} that  
 the lepton-flavor changing process 
 can induce the suitable magnitude of the muon anomalous magnetic 
 moment ($g_\mu -2$) in the decoupling solution 
 of the minimal supersymmetric standard model (MSSM) 
 satisfying the current experimental bound of the 
 branching ratio of $\tau \rightarrow 
 \mu \gamma$. 

In this paper we consider the 5 dimensional theory 
 where SUSY breaking brane is put on the center of 
 the right-handed electron of 
 the 2nd and the 3rd generation. 
 In this setup, the SUSY decoupling solution is obtained and 
 the sizable flavor mixings both in the left- and right-handed 
 sleptons can induce the suitable magnitude of $g_\mu -2$
 within the experimental bounds of 
 the lepton flavor violating processes.

\section{Decoupling in the extra dimension}

Let us start to 
 see how 
 the fermion mass hierarchies and SUSY decoupling solutions 
 come out \cite{HM3}.

Consider the up-type Yukawa coupling, 
 for example,
%
\bea
W = \int dy \delta(y) Q_i(x,y) \bar{U}_j(x,y) H_u(x), 
\eea
where $x$ denotes the coordinate of 
 four dimensional Minkowski space-time, 
 $y$ is a fifth spatial coordinate of five dimensions. 
$i,j$ are the generation indices. 
The order one coefficient is implicit. 
$Q_i$, $\bar{U}_i$ and $H_u$ are the chiral superfield  
 which transform as $({\bf 3}, {\bf 2}, 1/6)$, 
 $({\bf 3}^*, {\bf 1}, -2/3)$ and $({\bf 1}, {\bf 2}, 1/2)$
 under the Standard Model (SM) gauge groups, 
 $SU(3)_C \times SU(2)_L \times U(1)_Y$. 
We assume here that the MSSM matter fields 
 live in the bulk and 
 Higgs fields are localized on a brane at $y=0$. 
Integrating out the fifth dimensional degrees of freedom, 
 we obtain the effective Yukawa coupling in four dimensions 
 at the compactification scale as, 
%
\bea
\label{yukawa}
(y_{{\rm eff}})_{ij} \simeq {\rm exp}[-a^2 (y^2_{Q_i} 
+ y^2_{\bar{U}_j})], 
\eea
where we assume the form of the zero mode wave function of 
 the matter superfields to be Gaussian such as 
 ${\rm exp}[-a^2(y - y_{\Phi_i})^2]$, 
 where 
 $a$ is the inverse width of the zero mode wave functions, 
 $y_{\Phi_i}$ is the coordinate where the matter superfield 
 $\Phi_i(=Q_i, \bar{U}_i, \bar{D}_i, ...)$ is localized. 
As is clear from (\ref{yukawa}), 
 the information of Yukawa hierarchy is interpreted as 
 the ``geography" of configuration of the matter fields 
 in the extra dimensions. 
In Ref.\cite{HM3}, 
 we have found various types of fermion mass matrices 
 well describing the fermion mass hierarchies and 
 the flavor mixings in the fat brane scenario. 

We note that 
 the sfermion masses correlate with the fermion masses 
 in extra dimensions 
 by introducing SUSY breaking brane 
 because the sfermion masses are determined by the overlap of 
 wave functions of the matter fermions and the chiral superfields 
 with nonvanishing F-term VEV on SUSY breaking brane. 
Now that we know various matter configurations consistent 
 with experimental data, 
 the sfermion mass spectrum can be calculated and predicted. 

Let us discuss the sfermion mass in more detail. 
In our previous paper, 
 we have proposed the mechanism to generate the sfermion masses 
 in the fat brane scenario \cite{HM2}. 
``SUSY breaking brane", is introduced at $y=L$, 
 where the chiral superfields $X$ with nonvanishing F-term VEV 
 ($X=\theta^2 F$) is assumed to be localized. 
The extra vector-like superfields $\Phi'$, $\bar{\Phi}'$ 
 with mass $M < M_*$, 
 where $M_*$ is the five dimensional Planck scale, 
 are also introduced and 
 assumed to be localized on a SUSY breaking brane. 
We consider here the following superpotential, 
\bea
\label{sp}
W &=& \int dy \delta(y-L) [\frac{\lm}{\sqrt{M_*}} X(x) 
\Phi_i(x,y) \bar{\Phi}'(x) 
+ M \Phi(x)' \bar{\Phi}'(x)], \\
&=& \frac{\lm}{\sqrt{M_* a^{-1}}} 
{\rm exp}[-a^2(L-y_{\Phi_i})^2] X(x) \Phi_i(x) 
\bar{\Phi}'(x) + M \Phi'(x) \bar{\Phi}'(x), 
\eea
where $\lm$ is a dimensionless constant of order unity. 
Below the scale $M$, 
 we can integrate out the massive superfields $\Phi'$ and 
 $\bar{\Phi}'$, then the superpotential vanishes and 
 the effective K\"ahler potential are generated 
 at tree level, 
\bea
\label{effK}
\delta K_{{\rm eff}} = \frac{1}{\sqrt{M_* a^{-1}}} 
\frac{1}{M^2} {\rm exp}
[-a^2 \{ (L-y_{\Phi_i})^2 + (L-y_{\Phi_j})^2 \}] 
X^\dag X \Phi_i^\dag \Phi_j. 
\eea
The sfermion masses coming from (\ref{effK}) 
 at the compactification scale are 
\bea
\label{sfermion}
\tilde{m}^2_{ij} \simeq \frac{1}{\sqrt{M_* a^{-1}}} {\rm exp}
[-a^2 \{ (L-y_{\Phi_i})^2 + (L-y_{\Phi_j})^2 \}] 
\frac{|F|^2}{M^2}.  
\eea
It is crucial that the scale suppressing the K\"ahler potential 
 is replaced with $M < M_*$ not so as to be negligibly 
 small.\footnote{Without 
 introducing the extra vector-like superfields, the sfermion masses are 
 negligibly small due to the exponential suppression \cite{HM2}.} 
Note that $F<M^2$ is assumed in this argument and 
 also the overall sign of the K\"ahler potential is assumed 
 to be positive. 

We would like to mention the gaugino mass in our scenario. 
The gaugino masses are generated at tree level 
 since we assume that the gauge supermultiplets live 
 in a thick wall, 
%
\bea
\label{gmass}
\delta(y-L) \int d^2 \theta \frac{X(x)}{M_*^2} 
W^{\alpha}(x,y) W_{\alpha}(x,y) 
\Rightarrow M_{\lambda} = \frac{F}{M_*^2 L_c}, 
\eea
where $W_\alpha$ is the field strength tensor superfield and 
 $L_c$ is the width of the thick wall which should be considered as 
 the compactification length in our framework. 
For the gaugino masses to be around 100 GeV, 
 so we obtain 
\bea
\label{gaugino}
\frac{F}{M_*} \simeq 100(M_* L_c). 
\eea
%
A-terms are also induced only if the SUSY breaking brane 
 is located at the point where Higgses are located \cite{HM3}.

Now we consider whether 
 the sizable flavor mixings in the left- and right- handed slepton
 mass matrices can be obtained in the decoupling solutions. 
The decoupling spectrum of the slepton masses in five dimensional
 theory has been disscussed in Ref.\cite{HM3}. 
However, their flavor mixings have not been discussed, 
 which is the issue to be addressed in this paper. 

The strategy is the following. 
In order to obtain the decoupling spectrum and the large mixing 
 between the 2nd and the 3rd generations 
 both in the left- and right-handed sleptons simultaneously, 
 we set SUSY breaking brane at the point where
 the distances from $\bar{E}_{2,3}$ are the same.

In this setup, 
 the slepton mass matrices becomes as 
\bea
\tilde{m}^2_{ij} \sim {\rm exp}[-a^2 
\{(s-y_{\Phi_i})^2 + (s -y_{\Phi_j})^2 \}] 
\frac{F^2}{M^2}, 
\eea
where $s=(y_{\bar{E}_2} + y_{\bar{E}_3})/2$. 
We will see explicitly the slepton mass matrices for 
 the ``anarchy type'' fermion mass matrix 
 and improvement I, II \cite{HM3}. 
On the other hand, 
 we do not introduce extra vector-like superfields and 
 the relevant superpotential (\ref{sp}) for the quark sector. 
Therefore, squark masses are negligibly small, 
 radiatively induced at the weak scale by the gaugino RGE effects 
 and flavor blind.

\section{Sfermion mass matrices}

The ``anarchy type'' fermion mass matrices with large $\tan \beta$
 are obtained when the matters are localized in a five 
 dimensional coordinate as \cite{HM3}\footnote{These coordinates are 
 in units of $a^{-1}$.}
\bea
\label{anarchy2}
\begin{array}{lll}
y_{Q_1}^2 \simeq -2\ln \ep, & y_{\bar{U}_1}^2 \simeq -2\ln \ep, 
& y_{\bar{D}_1}^2 \simeq 0, \\
y_{Q_2}^2 \simeq -\ln \ep, & y_{\bar{U}_2}^2 \simeq -\ln \ep, 
& y_{\bar{D}_2}^2 \simeq 0, \\
y_{Q_3}^2 \simeq 0, & y_{\bar{U}_3}^3 \simeq 0, 
& y_{\bar{D}_3}^2 \simeq 0, \\
y_{L_1}^2 \simeq 0, & y_{\bar{E}_1}^2 \simeq -2\ln \ep, 
& y_{\bar{N}_1}^2 \simeq 0, \\
y_{L_2}^2 \simeq 0, & y_{\bar{E}_2}^2 \simeq -\ln \ep, 
& y_{\bar{N}_2}^2 \simeq 0, \\
y_{L_3}^2 \simeq 0, & y_{\bar{E}_3}^2 \simeq 0, 
& y_{\bar{N}_3}^2 \simeq 0. 
\end{array}
\eea
%
We set the parameter $\ep$ to be of order $\lm^2$, and 
 $\lm$ is the Cabibbo angle, $\lm \simeq 0.2$. 
The configuration (\ref{anarchy2}) generates 
 the following mass matrices for up, down quark sectors 
 and the charged lepton sector:  
\bea
\label{anarchy4}
m_u \simeq 
\left(
\begin{array}{ccc}
\ep^4 & \ep^3 & \ep^2 \\
\ep^3 & \ep^2 & \ep \\
\ep^2 & \ep & 1
\end{array}
\right)\langle H_u \rangle,~
m_d \simeq 
\left(
\begin{array}{ccc}
\ep^2 & \ep^2 & \ep^2 \\
\ep & \ep & \ep \\
1 & 1 & 1 \\
\end{array}
\right) \langle H_d \rangle, 
m_l \simeq 
\left(
\begin{array}{ccc}
\ep^2 & \ep & 1 \\
\ep^2 & \ep & 1 \\
\ep^2 & \ep & 1
\end{array}
\right) \langle H_d \rangle.  
\eea
On the other hand, 
 there are no mass hierarchies 
 in the neutrino mass matrix 
 since both left- and right-handed neutrinos
 are localized at the same point. 
The light neutrino mass matrix ($m_\nu^{(l)}$)
 through the see-saw mechanism\cite{seesaw} is given by 
\bea
m_\nu^{(l)} \simeq \frac{m_\nu^D (m_\nu^D)^t}{m_N} 
\simeq \left(
\begin{array}{ccc}
1 & 1 & 1 \\
1 & 1 & 1 \\
1 & 1 & 1 \\
\end{array}
\right)
\frac{\langle H_u \rangle^2}{M_R}. 
\eea
where $M_R$ is around $10^{15-16}$ GeV.\footnote{$M_R \simeq 10^{15-16}
$ 
 GeV is naturally obtained from the VEV of 
 the singlet field \cite{HM3}.} 
All elements of the above matrices 
 have $O(1)$ coefficients. 
These fermion mass matrices can 
 naturally explain 
 why the flavor mixing in the quark sector 
 is small while that in the lepton sector is large 
 \cite{Babu,various,HaMu}. 
The above fermion mass hierarchies and flavor mixings 
 are roughly consistent with the experimental data, 
 and explicit values of $O(1)$ 
 coefficients of mass matrices can really induce 
 the suitable magnitudes of fermion masses and 
 flavor mixing angles \cite{Babu}.

Here let us show one explicit example of coefficients. 
According to the method of determinig coefficients in 
 Ref.\cite{Babu}, 
 we suggest fermion mass matrices with ${\cal O}(1)$ coefficients as 
\begin{equation}
m_U=\left(
\begin{array}{ccc}
0 & 2d\epsilon ^3 & 0 \\
2d\epsilon ^3 & \frac{4}{5}c \epsilon ^2 & 0\\
0 & b\epsilon & 1\\
\end{array}
\right), \;
m_D=\left(
\begin{array}{ccc}
d\epsilon ^2 & d\epsilon ^2 & d\epsilon ^2 
\\
-d\epsilon  & d\epsilon  & d\epsilon \\
 \frac{c}{2}& b & 1\\
\end{array}
\right) ,
\end{equation}
\begin{equation}
m_L=\left(
\begin{array}{ccc}
\epsilon ^2 & 0  & 0 \\
b\epsilon ^2 & -2c\epsilon  & 0 \\
 0& -b\epsilon & 5\\
\end{array}
\right), \;
m_\nu^{(l)} =\left(
\begin{array}{ccc}
e & e & 0 \\
e & c  & 2.5\\
 0& 2.5 & 5\\
\end{array}
\right) .
\end{equation}
Here we take $b=4$, $c=3.6$, $d=2$, and 
$e=1.0$, then the CKM \cite{CKM} and the MNS \cite{MNS} 
 matrices are given by 
\begin{equation}
V_{CKM}=U_u^\dagger U_{d}=\left(
\begin{array}{ccc}
0.9984   & -0.05650   & -0.000373   \\
0.05649   & 0.9983   & -0.01197   \\
0.001049   & 0.01193   & 0.9999  \\
\end{array}
\right) ,
\end{equation}
\begin{equation}
U_{MNS}=U_l^\dagger U_{\nu }=\left(
\begin{array}{ccc}
0.8417   & -0.5322   & 0.0903  \\
-0.4717   & -0.6437   & 0.6025   \\
-0.2625   & -0.5498   & -0.7929  \\
\end{array}
\right) ,
\end{equation}
respectively. 
Where $U_u,U_d,U_l$, and $U_{\nu }$ are 
 defined as 
 $(U_u)^\dagger m_Um_U^\dagger U_u=(m^2_{U})_{diagonal}$, 
 $(U_d)^\dagger m_Dm_D^\dagger U_d=(m^2_{D})_{diagonal}$,
 $(U_l)^\dagger m_Lm_L^\dagger U_l=(m^2_{L})_{diagonal}$, and 
 $(U_\nu )^\dagger m_\nu m_\nu ^\dagger U_{\nu }=
 (m^2_{\nu })_{diagonal}$, respectively. 
The fermion mass hierarchies are given by 
\begin{equation}
m_t:m_c:m_u=1.012   : 0.0045    : 0.000014   , \;\;\;\;\;
m_b:m_s:m_d=4.49    : 0.12     : 0.00027    , 
\end{equation}
\begin{equation}
m_\tau :m_\mu : m_e=5.00   :0.28    :0.0015   ,  \;\;\;\;\;
m_{\nu_\tau}: m_{\nu_{\mu}} : m_{\nu_e}= 6.92    : 1.93    : 0.87    ,
\end{equation}
%
%
%
They are consistent with today's experimental results. 
The neutrino mass spectrum is suitable for 
 the LMA solar solution.

Now we calculate 
 the slepton mass matrices. 
For the decoupling solution, 
 we consider the case that 
 the SUSY breaking brane is located at the point, $y=s$. 
Then 
 the slepton mass matrices take the form as  
\bea
&&\tilde{M}^2_{L} 
\simeq \ep^{1/2} 
\left(
\begin{array}{ccc}
1 & 1 & 1 \\
1 & 1 & 1 \\
1 & 1 & 1
\end{array}
\right) \left( \frac{F}{M} \right)^2,\\
&&
\tilde{M}^2_E \simeq \ep^{1/2} 
\left(
\begin{array}{ccc}
\ep^{4 - 2\sqrt{2}} & \ep^{2 - \sqrt{2}} & \ep^{2 - \sqrt{2}} \\
\ep^{2 - \sqrt{2}} & 1 & 1 \\
\ep^{2 - \sqrt{2}} & 1 & 1 \\
\end{array}
\right) \left( \frac{F}{M} \right)^2. 
\eea
In the wide range of parameter regions of 
 order one coefficients, the 3rd generation 
 sfermions become light. 
The case of the rank of mass matrices being reduced to be 2 
 is the typical example. 
This situation is realized when $\lambda$ and $M$ in Eq.(3) 
 are common for the 2nd and 3rd generations. 
In this case the masses of the 1st and the 2nd generations 
 are heavy enough, which is just 
 the realization of decoupling solution. 
On the other hand, 
 the masses of the 3rd generation can be at least of order 100 GeV 
 through the gravity mediated effects, which always generate 
 $O(100)$GeV sfermion masses. 
This gravity effects will be explained later. 
As for the right-handed slepton, 
 the 1st generation masses are of order $0.15$ times
 smaller than those of the 2nd and 3rd generations as 
 in Eq.(20). 
However there are parameter regions 
 where the 3rd generation sfermions become light 
 as in the above situation.

Here we show one example of the slepton mass matrix 
 with ${\cal O}(1)$ coefficients, which 
 induces the decoupling solution. 
Denoting $6 \times 6$ sfermion mass matrix as 
\bea
\tilde{M}^2 = \left(
\begin{array}{cc}
\tilde{M}^2_{L} & \tilde{M}^2_{LR} \\
\tilde{M}^2_{RL} & \tilde{M}^2_{R} \\
\end{array}
\right), 
\eea
we suggest 
%
\begin{equation}
\tilde{M}^2_L\simeq \epsilon^{\frac{1}{2}}
\left(
\begin{array}{ccc}
2 & 1 & 1\\
1 & 5.7 & 5.64\\
1   & 5.64 & 5.6\\
\end{array}
\right) \left(\frac{F}{M} \right)^2, 
\end{equation}
\begin{equation}
\tilde{M}^2_E \simeq \epsilon^{\frac{1}{2}} 
\left(
\begin{array}{ccc}
3 \times \epsilon ^{4-2\sqrt 2} & 
- \epsilon ^{2-\sqrt 2} & 
 -\epsilon ^{2-\sqrt 2}\\
-  \epsilon ^{2-\sqrt 2 } & 5.7  & 5.64 \\
- \epsilon ^{2-\sqrt 2}      & 5.64 & 5.6 \\
\end{array}
\right) \left( \frac{F}{M} \right)^2.
\end{equation}
The mass eigenvalues of $\tilde{M}^2_L$ and $\tilde{M}^2_E$ 
 are given by 
\begin{equation}
\tilde{m}_{e_L} : \tilde{m}_{\mu_L} : \tilde{m}_{\tau_L} \simeq 
 6.69 \:{\rm TeV} : 17.0 \:{\rm TeV}: 379 \:{\rm GeV}, 
\end{equation}
\begin{equation}
\tilde{m}_{e_R} : \tilde{m}_{\mu_R} : \tilde{m}_{\tau_R} \simeq 
  1.27\:{\rm TeV} :  16.8\:{\rm TeV}: 380 \:{\rm GeV}, 
\end{equation}
where $F/M \simeq \sqrt{5} \times 5$ TeV. 
We diagonalize these mass matrices in the basis 
 which the fermion mass matrices are diagonal. 
In this basis $\tilde{M}^2_L$ and $\tilde{M}^2_E$ 
 are changed as 
\begin{equation}
M^2_L=U^\dagger _l \tilde{M}^2 _LU_l, \;\;\;
M^2_E=U_r^\dagger \tilde{M}^2_E U_r, 
\end{equation}
where $U_r$ is defined as 
 $U_l^{\dagger } m_L U_r=(m_L)_{diagonal}$. 
$M^2_L$ is diagonalized by the unitary matrix 
\begin{equation}
V_L =\left(
\begin{array}{ccc}
0.9891 & 0.14708  & 0.0033   \\
-0.106  & 0.7011 & 0.70496 \\
0.1013 & -0.6976  & 0.7092\\
\end{array}
\right) .
\end{equation}
This means that 
 the mixing angles are 
\begin{equation}
\label{30}
\sin \theta _{12} = 0.14708, \;\;\; \sin \theta _{13} = 0.0033, \;\;\;
 \hspace{0.2cm} \sin \theta _{23}=0.70496 .
\end{equation}
$M^2_E$ is diagonalized by the unitary 
 matrix 
\begin{equation}
V_R=\left(
\begin{array}{ccc}
0.9999 & -0.002605   & 0.00072     \\
0.00240   & 0.7325 & -0.680 \\
0.00124   & 0.6806 & 0.7325\\
\end{array}
\right).
\end{equation}
This means that 
 the mixing angles are 
\begin{equation}
\label{32}
 \sin \theta _{12} = -0.002605, \;\;\; \sin \theta _{13} = 0.00072, \;
\;\;
 \sin \theta _{23} = -0.680.
\end{equation} 
Above one example of $O(1)$ coefficients 
 really induce $O(1)$ mixings between the 2nd and the 
 3rd generations in both left- and right-handed 
 slepton sectors, and also 
 light 3rd generation sfermion masses. 
There are wide parameter region where 
 the same situation is satisfied. 
In Ref.\cite{CHH} it has been shown that 
 this parameter region can induce  
 enough large muon anomalous magnetic moment, 
 $g_{\mu}-2$, 
 within the constraint of $\tau \rightarrow \mu \gamma$.  
The severer constraint exists in 
 the process of $\mu \rightarrow e \gamma$. 
However it can be satisfied since 
 the mixing angles between the 1st and the 3rd generations 
 can be of $O(10^{-3})$ in the wide parameter region 
 as in Eqs.(\ref{30}) and (\ref{32}).

The improved mass matrices I (large tan$\beta$)
 in Ref.~\cite{HM3} induce the same sfermion mass matrices 
 in Eqs.(22) and (23). 
Thus, this case also realizes the suitable 
 decoupling solution. 
On the other hand, the matter configurations of 
 ``anarchy type'' with small tan$\beta$ and 
 improved I with small tan$\beta$ cannot 
 induce the decoupling solutions. 
It is because the smallest 
 sfermion mass eigenvalues are 
 given by 
 $M^2_{L} \simeq \ep^{3/2+\sqrt{2}}(F/M)^2$ and 
 $M^2_{E} \simeq \ep^{3/2-\sqrt{2}}(F/M)^2$ in 
 these cases. 
They are different in more than two order
 magnitude from each other,  
 which are not suitable for the decoupling solutions.

As for the improved mass matrices II in Ref.~\cite{HM3}, 
 it is not suitable in our scenario as the following argument.  
Note that ${\cal O}(100{\rm GeV})$ sfermion mass of 
 the 3rd generation is coming from the gravity mediation 
 as mentioned above, 
\bea
\delta(y-s)\int d^4\theta \frac{X^\dag X}{M_*^3}Q_i^\dag Q_j 
\to \tilde{m}_{ij}({\rm gravity}) \sim \frac{F^2 a}{M_*^3} 
{\rm exp}[-(s-y_{Q_i})^2-(s-y_{Q_j})^2]. 
\eea
The gravity mediated scalar masses are related to 
 the gaugino mass as 
\bea
\tilde{m}_{ij}({\rm gravity}) \sim 
\left\{ 
\begin{array}{l}
0.10(M_*L_c)(aL_c)M_\lambda^2~({\rm Anarchy,Improvement I}), \\
0.76(M_*L_c)(aL_c)M_\lambda^2~({\rm Improvement II}). 
\end{array}
\right.
\eea
For $\tilde{m}_{ij}({\rm gravity})$ to be of order 100 GeV, 
\bea
(M_*L_c)^2 \simeq 
\left\{ 
\begin{array}{l}
10~({\rm Anarchy,Improvement I}), \\
1~({\rm Improvement II}) 
\end{array}
\right.
\eea
are obtained if $a \simeq M_*$ for simplicity. 
For example, 
 $M_* \simeq \sqrt{10}\times 10^{16}$ GeV, $L_c^{-1} \simeq 10^{16}$ GeV is viable 
 for Anarchy, Improvement I. 
However, Improvement II seems to be unnatural 
 since $M_*L_c \simeq 1$ contradicts the constraint 
 $L^{-1}< a \le M_*$ in the fat brane scenario.

\section{Summary}

We have discussed the fermion mass hierarchy and the flavor mixings 
 in the fat brane scenario of a five dimensional SUSY theory 
taking into account of ${\cal O}(1)$ coefficients. 
We consider the case 
 where SUSY breaking brane is put on the center of 
 the 2nd and the 3rd generations' right-handed electron fields 
 in the 5 dimensional coordinate. 
In this case the decoupling solution is realized. 
The sizable flavor mixings both in the left- and right-handed 
 sleptons can be naturally induced, 
 which can realize the suitable magnitude of 
 the muon anomalous magnetic moment 
 within the experimental bounds of lepton flavor violating processes.

%
     %
%

\begin{flushleft}
{\Large\bf Acknowledgments}
\end{flushleft}
We thank J. Hisano and G-C. Cho 
 for many helpful discussions. 
N.H. is supported by the Grant-in-Aid for Scientific Research, 
Ministry of Education, Science and Culture, Japan 
(No.14039207, No.14046208, No.14740164) and 
N.M. is supported 
by the Japan Society for the Promotion of Science 
for Young Scientists (No.08557).

\vspace{1cm}
%
     %
%

\end{document}